\documentclass[11pt,letterpaper]{llncs}
\usepackage[ruled]{algorithm2e}
\usepackage{amsmath}
\usepackage{amssymb}
\usepackage{graphicx}
\usepackage{marvosym}
\usepackage{url}
\usepackage{fancyhdr}
\usepackage{float}
\usepackage{geometry}
\newcommand{\keywords}[1]{\par\addvspace\baselineskip
	\noindent\keywordname\enspace\ignorespaces#1}

\begin{document}
	
	\title{ENHANCING NETWORK FORENSICS with PARTICLE SWARM and DEEP LEARNING: THE PARTICLE DEEP FRAMEWORK}
	\author{Nickolaos Koroniotis \inst{1} \and Nour Moustafa\inst{1}}
	\institute{School of Engineering and Information Technology, University of New South Wales Canberra, Canberra, Australia\\\email{n.koroniotis@student.adfa.edu.au}}
	\maketitle
	\begin{abstract}
		The popularity of IoT smart things is rising, due to the automation they provide and its effects on productivity. However, it has been proven that IoT devices are vulnerable to both well established and new IoT-specific attack vectors. In this paper, we propose the Particle Deep Framework, a new network forensic framework for IoT networks that utilised Particle Swarm Optimisation to tune the hyperparameters of a deep MLP model and improve its performance. The PDF is trained and validated using Bot-IoT dataset, a contemporary network-traffic dataset that combines normal IoT and non-IoT traffic, with well known botnet-related attacks. Through experimentation, we show that the performance of a deep MLP model is vastly improved, achieving an accuracy of 99.9\% and false alarm rate of close to 0\%.
		\keywords{Network forensics, Particle swarm optimization, Deep Learning, IoT, Botnets}
	\end{abstract}

	\section{Introduction}
	With more than 7 billion devices deployed in 2018 and double that number in 2019, 
	smart IoT things are becoming ever more popular, as they provide automated services 
	that improve performance and productivity, while reducing operating costs \cite{Cook2019}. 
	Various applications of the IoT have been developed, such as the smart home comprised of 
	home appliances like smart lights and fridges, and on a larger scale the Industrial IoT (IIoT) 
	that spans areas  such as industry, healthcare, agriculture and automation. The next step 
	is considered to be the ''smart city'', where smart things will be used to monitor and 
	control powerplants, public transport, water supply and more \cite{Cook2019}.
	
	However, it has been proven that IoT devices are vulnerable to both well 
	established and new IoT-specific attack vectors. In a 2018 report by Symantec regarding the 
	security threats found in the Internet \cite{Corporation2019}, it was reported that 
	the total number of attacks targeting IoT devices for 2018 exceeded 57,000, with 
	more than 5,000 attacks being recorded each month. Hackers have compromised 
	vulnerable, unpatched or unencrypted IoT devices in order to steal sensitive data, 
	corrupt the device's normal operation, spread malware infections \cite{koroniotis2019forensics}\cite{ali2018cyber}\cite{Corporation2019} or even compromise 
	the security of a smart home by disabling smart locks and garage doors \cite{Robberts2019}.

	Due to the lack of common standards and heterogeneity displayed by the IoT 
	\cite{koroniotis2019forensics}, the development of an efficient network forensic solution 
	becomes difficult \cite{conti2018internet}, as there may be multiple diverse protocols in use in a 
	single deployment \cite{ronen2017iot,meffert2017forensic,al2017internet}. Typically, the network forensic process is segmented into several
	distinct phases, whereby each phase defines the necessary preparation,
	analysis and actions of investigation \cite{kaur2012digital}, with these phases being 
	identification, collection, preservation, examination, analysis and presentation. The first three 
	phases define access to the crime scene, detection of potential sources of evidence, secure 
	collection of digital artifacts and traces and their preservation. Examination and analysis 
	define the actions necessary to identify useful evidence in the collected data, and then inferences 
	are made about the crime, while the presentation stage prepares the identified information to be 
	presented in court of law.
	
	Many forensic frameworks that rely on the public ledger scheme have been developed, 
	although they focus on the acquisition stage rather than the entire investigation phase 
	\cite{meffert2017forensic} \cite{hossain2018fif} \cite{hossain2018probe} \cite{Cebe2018}. 
	By relying on a public ledger scheme, the integrity of evidence would be guarantied, while 
	experts would have immediate access to it. These frameworks however, do not cover the 
	examination and analysis stages, and may introduce some disadvantages to the forensic process
	such as privacy concerns, as the user's data is distributed between stakeholders while adding 
	excess complexity 
	
	During a forensic investigation, some aspect of a computer system is 
	examined to identify traces. One source that is preferred for IoT investigations, 
	is traffic collection, with two main approaches: deep packet inspection and network flow
	analysis. Deep packet inspection focuses on the payload allowing for an in-depth analysis. 
	Network flow analysis summarizes network traffic, by ignoring the payload and utilising 
	statistical data like connection speed and exchanged bits. For the requirements of the 
	research presented in this paper, network flow is preferred, as we suggest in this work 
	\cite{koroniotis2019forensics}.
	
	Since IoT devices are designed to be continuously active, 
	collecting traffic from such a network would result in excessive 
	volumes of data. As such, to analyse the collected data, fast and 
	automated methods are employed, with one prominent example 
	being deep learning that is ideal for rapidly scanning large volumes 
	of network data to detect patterns that indicate and attack 
	\cite{shone2018deep,prabakaran2018survey,wang2015applications,zhao2017intrusion}.
	However, in order to utilise a deep learning model, it first needs to be trained, which 
	involves selecting hyperparameters, the values of which can greatly affect its performance 
	\cite{Zela2018,wang2018stealing}. Thus, researchers have sought to develop ways of selecting
	optimal values for a model's hyperparameters \cite{chen2018learning,Zela2018,wang2018combination,stamoulis2018hyperpower}, with an emphasis given towards 
	more automated processes. Regardless, no single optimization method has been accepted 
	as the preferred method by the research community and as such, the research is ongoing. 
	It is a necessity to develop network forensic-based optimisation to timely investigate security 
	incidents in IoT networks \cite{watson2016digital,koroniotis2019forensics,chernyshev2018internet} .

	The main contributions of this paper are as follows:
	\begin{itemize}
		\item {We propose a new network forensic framework, named Particle
			Deep Framework (PDF), based on optimisation and deep learning.}
		\item {We use an optimization method based on Particle Swarm Optimization(PSO)
			to select the hyperparameters of the Deep Neural Network (DNN).}
	\end{itemize}

The structure of the paper is as follows. Section 2 includes background and related
research in the application of particle swarm optimisation and deep neural networks to network forensics.
Section 3 presents our proposed Particle Deep Framework in detail. Section 4 presents and discusses the experimental results acquired by utilising our PDF.
In Section 5 we discuss the advantages and disadvantages of the PDF.
 Finally, Section 6 includes the concluding remarks of this paper.
	
	\section{Background and Related Work}
	\subsection{Digital Forensics}
	Due to the malicious actions of hackers who take advantage of 
	vulnerabilities present in popular technologies, the field of digital forensics emerged.
	Through digital forensics, experts examine a crime scene, gather data that are then processed
	and analysed in order to identify a hacker's methods and target, a process that can lead to 
	prosecution \cite{palmer2001road}. Since it was first introduced, digital forensics has been 
	separated into specialized sub-disciplines, each ideal for different areas, such as: cloud forensics, IoT forensics, network forensics and data forensics \cite{joseph2019analysis}.
	
	Network forensics, which is the subdiscipline that we utilise in this paper, 
	examines security network-related security incidents, with collected data spanning
	from logs to packet captured. Practitioners of network forensics often employ
	automated software and hardware tools for the collection and preservation 
	of data, however, the process of performing a forensic examination is not rigidly defined.
	This has resulted in the emergence of various digital forensic frameworks, which determine 
	the correct course of action during an investigation, separating the process into autonomous 
	stages and suggesting appropriate tools and techniques for each task. Even though many 
	forensic frameworks have been proposed \cite{hossain2018probe,le2018biff,hossain2018fif}, 
	existing solutions for smart homes give emphasis on acquisition and neglect examination 
	and analysis. Furthermore, no single framework has been acknowledged as superior, due 
	to lack of standardization in IoT and heterogeneity of computer systems 
	\cite{valjarevic2015comprehensive,caviglione2017future,koroniotis2019forensics}.
	
	Various researches have proposed forensic frameworks for IoT environments 
	\cite{meffert2017forensic} \cite{hossain2018fif} \cite{hossain2018probe} \cite{Cebe2018}.
	Hossain et al cite{hossain2018probe} proposed Probe-IoT, an acquisition framework that 
	establishes chain of custody, while ensuring that privacy is maintained. An acquisition model 
	for smart vehicles named Block4Forensic was proposed by Cebe et al. \cite{Cebe2018}. Both Probe-IoT and Block4Forensic utilise a blockchain to ensure the integrity of the collected data, which includes diagnostic and interaction data acquired from IoT devices. 
	
	Most of the proposed IoT frameworks were constructed by using distributed blockchains 
	\cite{le2018biff,hossain2018fif,hossain2018probe}. Hossain et al. \cite{hossain2018probe} 
	proposed Probe-IoT and \cite{hossain2018fif} FIF-IoT, Le et al. \cite{le2018biff} developed BIFF. 
	These frameworks utilise a distributed blockchain, managed by several entities like the manufacturers, 
	the police and insurance companies, with pre-approved roles and digital signatures ensuring 
	confidentiality and non-repudiation. For the issuing of digital keys necessary for the digital signatures,
	a certification authority needs to be established, that will also manage the public-keys and avoid main-in-the-middle attacks \cite{han2016survey}.
	
	While these acquisition frameworks may improve an investigation, by providing easy access 
	to collected data, they have some drawbacks. To begin with, for the frameworks to
	function effectively, law enforcement need to invest in resources to store and manage
	the collected data, multiple independent organisations need to collaborate seamlessly
	while owners of IoT devices need to agree and trust the organisations that may access
	their data \cite{hossain2018probe,le2018biff,hossain2018fif}. In addition, owners may
	face extra charges due to the introduction of dedicated devices for data collection, or
	deterioration of their device's performance and increases in power consumption due
	to transmissions to services that incorporate the collected data to the blockchain.

\subsection{Deep Learning for tracing and discovering threat behaviours}

Commonly, network forensic applications have incorporated a number of techniques based 
on mathematics and machine learning, such as fuzzy logic, na\"ive bayes classifiers, 
support vector machines and neural networks \cite{liao2009network,ahmed2018sairf,yudhana2018ddos,nguyen2014approach}.
However, contemporary research has proposed deep learning as an alternative as, 
long training times notwithstanding, deep models tend to outperform other solutions 
when tasked with processing large volumes of data \cite{alrawashdeh2016toward,azmoodeh2018robust,koroniotis2019forensics,shone2018deep,zhao2017intrusion}.

Consisting of both discriminative and generative models, deep learning is a subgroup
of neural networks, that is designed to incorporate multiple hidden layers and neurons in
what is known as a "deep architecture" \cite{lecun2015deep,shone2018deep}. By stacking 
multiple (in the thousands) hidden layers, a deep learning model is capable of detecting more 
complex patterns, along with their variations, than simpler and shallower neural networks 
\cite{shone2018deep,Zela2018}.

As an application of network forensics, deep learning has been 
employed for attack detection in network traces, with multiple examples in research. 
Shone et al. \cite{shone2018deep} designed an intrusion detection system (IDS)
comprised of a pre-trained, non-symmetric deep autoencoder stacked with a random forest. 
The IDS, trained on the KDD dataset achieved an 89.22\% accuracy. For the purpose of 
malware detection, Azmoodeh et al. \cite{azmoodeh2018robust} developed a deep 
convolutional neural network, that was trained on eigenvectors obtained from 
smartphone application code. The model achieved 98\% precision and accuracy.

Brun et al. \cite{brun2018deep} investigated the detection of denial of service 
and sleep attacks targeting IoT gateways, and proposed a detection method 
based on a random neural network model although, its performance was similar 
to a threshold detector. A deep learning model based on MLP was proposed by 
Pektas et al. \cite{pektacs2018botnet} that focused on detecting botnets by 
flagging C\&C traffic. Results indicated that the performance of deep learning 
models, when tasked with identifying botnets in network flows, was acceptable. 

One of the first steps in utilising Convolutional Neural Networks (CNNs) for deep 
packet inspection, named D2PI, was proposed by Cheng et al. \cite{cheng2018d2pi}. 
This system was trained on extracted traffic payload data, with the CNN receiving a 
fixed input. Results reported by the researchers were promising. Alrashdi et al. 
\cite{alrashdi2019ad} developed AD-IoT, an anomaly NIDS designed 
to identify infected IoT devices. The NIDS utilised a random forest and extra tree 
classifiers and was evaluated on the UNSW-NB15 dataset, with results being promising.

Previous work on forensic frameworks for the IoT has either focused on the acquisition 
aspect of a forensic investigation \cite{hossain2018probe, le2018biff, hossain2018fif}, or 
requires alterations to the smartphone applications associated with a smart device 
\cite{babun2018iotdots}. As such, the framework proposed in this research, the PDF, is an 
important and practical addition to the network forensic research literature, as it incorporates 
both the examination and analysis phases by using deep learning and network flow data, 
without the need for alterations to IoT system architectures.

\subsection{The Particle Swarm Optimization algorithm}

In order to tune the hyperparameter values 
of a deep MLP model, in the context of this paper, we employed 
Particle Swarm Optimization (PSO). PSO is a metaheuristic swarm-based
optimisation algorithm developed by Eberheart and Kennedy in 1995 \cite{kennedy1995particle}.
At the start of PSO, a swarm of particles is randomly generated, and tasked with exploring
a variable search space, with each position being a new value for the variable that is being optimised. 
During each iteration, a particle's velocity and local position, as-well-as the global best position are updated, based on its previous position, its best-detected position (value) and the swarm's 
best position \cite{marini2015particle,wang2018particle}. The quality of a particle's location is 
determined by an objective function.

The original PSO algorithm proposed by Kennedy et al. \cite{kennedy1995particle}, 
which is explained by equations 3-6, was followed by a number of variants, each designed 
for different scenarios. For example, Kennedy et al. proposed in their original work, that the 
learning rates $\theta_1$ and $\theta_2$ in equation 5 greatly affect the search pattern of the 
swarm, with increases in $\theta_1$ prioritizing local search while increases in $\theta_2$ spreading 
the swarm. The next variant, called standard PSO, was proposed by Shi et al. \cite{shi1998modified} 
and integrated an inertia coefficient to equation 5 and specifically, multiplied it to the previous velocity 
value. As this inertia ($\omega$) has a direct effect on a particle's trajectory, various initialization 
strategies have been proposed, such as setting it to a positive, fixed value \cite{shi1998modified}, 
random initialization \cite{eberhart2001tracking} and using a function that declines with time 
cite{nikabadi2008particle }, with a linear example given in Equation 1. In Equation 1, $\omega_{max}$ 
and $\omega_{min}$ are user-defined maximum and minimum weights, while $i$ and $i_max$ indicate 
the the current and total number of iterations of the swarm respectively. The justification for using a
decaying inertia is to force particles to spread-search at early intervals of the PSO, and then gravitate 
towards identified optima.

\begin{equation} 
{\omega_t}={\omega_{max}}-{\frac{i}{i_{max}}*{\omega_{max}- \omega_{min}}} 
\end{equation}

Because experiments demonstrated the tendency of original and standard PSO to 
cause the velocity of particles to explode, improvements were suggested, 
in the form of velocity clamping and constriction factors \cite{eberhart1996computational,clerc1999swarm,eberhart2000comparing}.
Velocity clamping binds a particle's velocity to a pre-determined upper bound, 
while constriction factors alter Equation 5, by multiplying the new velocity with 
a constriction factor "K" given by Equation 2.

\begin{equation} 
K=2/|4-\phi-\sqrt{\phi^2-4\phi}|, where \phi=\theta_1+\theta_2 and \phi>4. 
\end{equation}

Other variants for the PSO algorithm were designed, to allow its application to 
new, previously not supported problems \cite{kennedy1997discrete}. Some prominent
examples include a binary version of PSO \cite{kennedy1997discrete}, where the restricted 
velocity of a particle was calculated and fed as input to a sigmoid function that produced a 
binary value (either '0' or '1'). The cooperative PSO \cite{van2004cooperative} was proposed 
for multi-dimensional problems, with a new swarm spawned for each dimension. Finally, the 
fully informed PSO \cite{mendes2004fully} alters Equation 5 and uses the best position of neighbours
to calculate a particle's new velocity.

	\section{The Particle Deep Framework}
\begin{figure}
		
		 \centering   \includegraphics[width=0.8\textwidth]{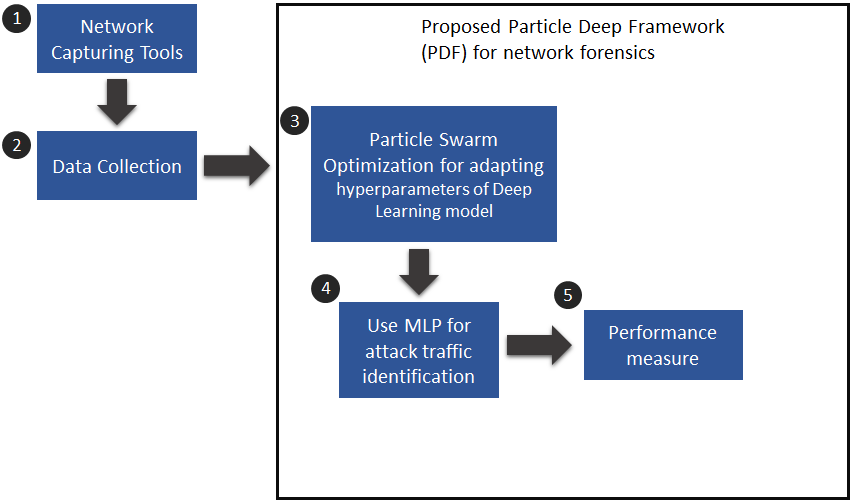}    
		
		\caption{\label{fig:PDF_Diagram} Proposed network forensic framework using
			particle swarm optimization and Multi-layer Perceptron (MLP) deep
			learning algorithms.}
	\end{figure}

	In this section, we introduce the Particle Deep Framework (PDF), a multi-staged novel network 
	forensic framework for detecting and analysing attacks and their origins in IoT networks, 
	that combines deep learning and particle swarm optimisation methods, as shown in Figure 
	\ref{fig:PDF_Diagram}.  
	
	Next, the five stages depicted in in Figure \ref{fig:PDF_Diagram} of the new PDF forensic framework are expanded as follows.
	
	\begin{itemize}
		\item \textbf{\large{}Stage 1: Network capturing Tools:}
		In this stage, IoT devices capable of accessing all local network traffic by 
		being set in promiscuous mode, are attached to a network that is under investigation. 
		Specialised packet capturing tools, such as Wireshark \cite{2018Wireshark}, 
		Tcpdump \cite{2018Tcpdump} and Ettercap \cite{2018Ettercap} are then employed
		to collect network packets. As an example, for the purposes of 
		developing the Bot-IoT dataset \cite{koroniotis2019towards}, we employed 
		T-shark \cite{2018p}, a terminal-based alternative to Wireshark. Through 
		the "-i" command we specified the Network Interface Card (NIC) to be used,
		which was set in promiscuous mode, allowing the collection of all the generated 
		packets in the local virtual network. The collected pcap files are then 
		utilised in the following data collection stage.
		
		\item \textbf{Stage 2: Data Collection and Management Tools:}
		
		In this stage, data collection takes place, producing results like the Bot-IoT and 
		UNSW-NB15 datasets. To ensure the preservation of the collected data, the digests 
		of the pcap files are generated by using an SHA-256 hashing function 
		\cite{armknecht2015privacy}. Network flows are then 
		generated from the collected pcaps, by using network flow extraction tools 
		like Argus \cite{2018j} or Bro \cite{2018a}. Further preprocessing actions like 
		feature normalization, elimination and extraction, improve the training process of 
		machine learning models at later stages. In the next stage, the produced data is 
		utilised by deep learning and particle swarm optimisation, to identify, trace and 
		analyse cyber-attacks.
		
		\item \textbf{Stage 3: Particle Swarm Optimization (PSO) for adapting
			hyperparameters of Deep Learning model:}
		
		In this stage, the hyperparameters of deep learning models are 
		tuned by an optimisation algorithm, with the PSO \cite{wang2018particle} 
		chosen for the task, due to its convergence speed compared to other 
		evolutionary algorithms \cite{marini2015particle,lu2015fuzzy}. In this study, 
		the PSO was used to identify hyperparameters that periodically maximise the Area Under 
		Curve (AUC) of a deep Multi-Layer Perceptron (MLP) model. The identified hyperparameters
		are then used to train the final version of the MLP model. 
		
		\item \textbf{Stage 4: MLP deep learning for attack identification:}
		
		In this stage, the preprocessed data from Stage 2 and the tuned 
		hyperparameters of Stage 3 are utilised to train and evaluate a deep MLP
		model. For this study, the deep MLP's architecture was comprised of seven layers
		and number of neurons as follows: 20, 40, 60, 80, 40, 10, 1. From stage 3, the 
		number of epochs, learning rate and batch size are optimised and used during 
		the training of the deep model, with the data from Stage 2 split into 80\% and 20\% 
		for training and testing respectively.
		
		\item \textbf{Stage 5: Performance measure:}

		In the final stage, the performance of the finalised deep MLP 
		model is gauged by parsing the testing set and obtaining the following metrics:
		accuracy, precision, recall, false positive and negative rates and F-measure. 
		In the following two subsections, Stages 3-5 are discussed in more detail.
	\end{itemize}
	
	 \subsection{Particle Swarm Optimization (PSO) Algorithm for deep learning parameter
		estimations}
	
	Originally inspired by observing the movement of animal swarms in their 
	natural habitat, Particle Swarm Optimisation (PSO) is a metaheuristic 
	evolutionary algorithm, that spawns a pre-determined number of particles ($P$) 
	that are randomly initialized and set to traverse a variable's search space ($v$).
	
	During a particle's propagation through the search space, each new position 
	(${v_{t+1}}$) is evaluated by the output of an objective function, which may change
	based on the problem being optimised. A particle is defined by four values (Equation 4),
	its velocity (${v_t }$) current position (${x_t}$), its local best (${x_{lbest}} $) and the
	swarm's best position (${x_{gbest}} $), as given in Equations 3,5,6.

	\begin{equation}    P={p_1,p_2,\dots,p_n},n \in \mathbb{N}   \end{equation} 
	\begin{equation}    \forall p_n \in P, p_n=({x_t}, {v_t }, {x_{lbest}}, {x_{gbest}})   \end{equation}
	 \begin{equation}     {v_{t+1}}= {v_t}+ \theta_1*rand*({x_{lbest}}-{x_t}) + \theta_2*rand*({x_{gbest}}-{x_t}), rand \in [0,1] \end{equation}   
	 \begin{equation}    {x_{t+1}}={x_t}+{v_{t+1}}   \end{equation}

	In Equation 5, ${v_{t+1}}$ a particle's new velocity is 
	determined by its previous velocity ${v_t}$, a random (rand) 
	proportion of the learning rates $\theta_1$, $\theta_2$ and 
	the differences (distance) of its current position to its local and 
	the swarm's best positions. Equation 6 gives the updated position 
	of a particle ${x_{t+1}}$, determined by its previous position ${x_t}$ and
	its current velocity. Next, Algorithm 1 depicts an iteration of the PSO algorithm
	used to maximise the AUC of a deep model \cite{wang2018particle}.
	
\begin{algorithm}[h]
	\label{alg:PSO_Max}
	\caption{Particle Swarm Optimization maximization algorithm}
	P $\leftarrow$ construct\_particles(n\_particles)\;
	$\forall$ p $\in$ P, $p.{X_{lbest}}=p.{x_0}, p.{X_{gbest}}=- \infty$\;
	epochs $\leftarrow$ load\_epochs()\;
	e$\leftarrow$0\;
	\While{e \textless epochs}{
		\ForEach{ p $\in$ P}{
			${v_{t+1}}= {v_t}+ \theta_1*rand*({x_{lbest}}-{x_t}) + \theta_2*rand*({x_{gbest}}-{x_t}), rand \in [0,1]$\;
			${x_{t+1}}={x_t}+{v_{t+1}}$\;
			\If{${x_{t+1}}$ \textgreater ${x_{lbest}}$}{
				${x_{lbest}}$=${x_{t+1}}$\;
			}
			\If{${x_{t+1}}$ \textgreater ${x_{gbest}}$}{
				${x_{gbest}}$=${x_{t+1}}$\;
			}}}
	\Return{P.global\_best()
	}

\end{algorithm}

	The PSO algorithm depicted in Algorithm \ref{alg:PSO_Max}, is
	set to maximise the AUC of a deep MLP model, in order to determine 
	the optimal values for the three hyperparameters: learning rate, 
	epochs and batch size, by separately traversing their search spaces. 
	
	There are a number of reasons that support the selection of PSO algorithm 
	in-place of another evolutionary metaheuristic algorithm for hyperparameter tuning. 
	To begin with, it has been established that PSO often converges faster than alternatives 
	\cite{Couceiro2016} and can produce acceptable results in realistic time \cite{bonyadi2017particle}. 
	Furthermore, its algorithm is readily understood and implemented \cite{ab2015comprehensive}. 
	Finally, this work provides empirical information about the performance of PSO for hyperparameter 
	tuning, in the context of deep learning and network forensics, as our research indicates this is 
	the first time it has been applied to the task.

	\subsection{Proposed Particle Deep Framework for Network Forensics}
\begin{figure}[H]    \centering    \includegraphics[width=0.8\textwidth]{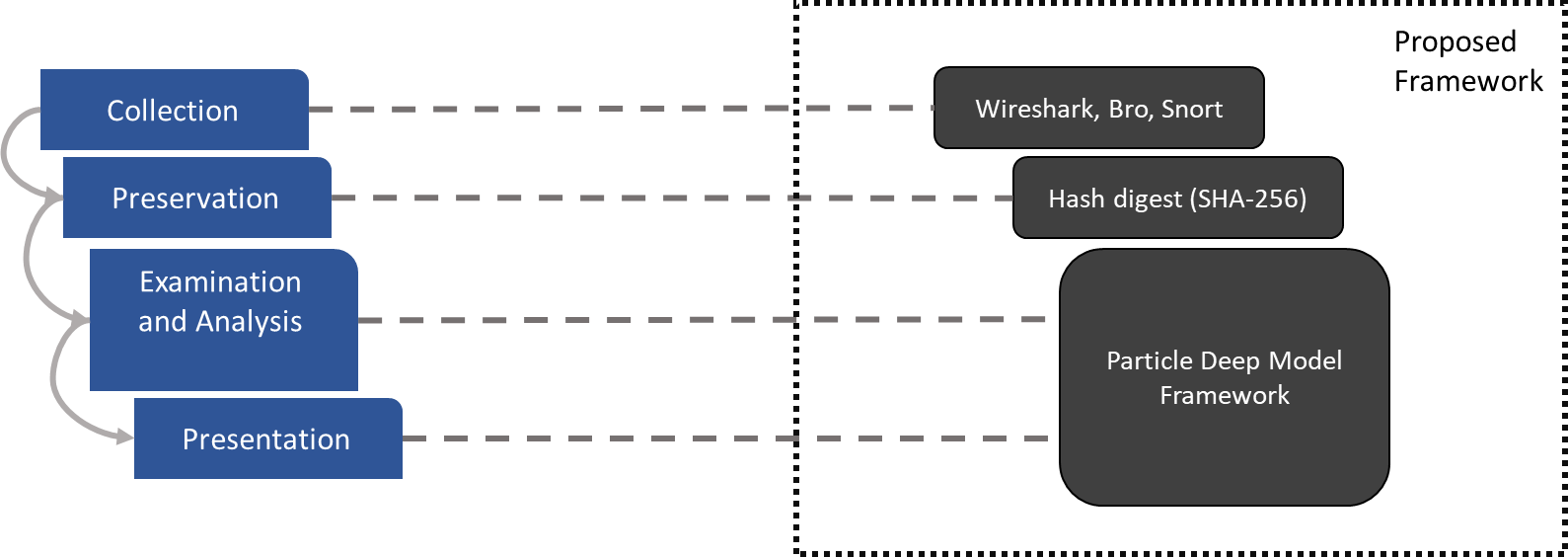}       
		
		\caption{\label{fig:Stages_of_investigation} Stages of investigation in network
			forensics including the proposed framework}
		
	\end{figure}

	The novel PDF is a considerable inclusion in the discipline of network forensics, 
	overlapping with the stages of network forensics, collection, preservation, 
	examination and analysis and presentation, as depicted in Figure \ref{fig:Stages_of_investigation}.
	The proposed framework takes advantage of a deep neural network's multiple layers, 
	which enhance the model's performance while maintaining the execution time within reason.
	
	\begin{algorithm} \caption{\label{alg:PDM_Opt} Particle deep model for hyperparameter estimation
			of deep learning}
		\KwData{}    nn $\leftarrow$ load\_neural\_network\_structure() \;    [b,e,lr] $\leftarrow$ initialize\_random\_hyperparameters() \;    hyperparameters  $\leftarrow$ [b,e,lr] \;    PS $\leftarrow$ construct\_particle\_swarm(n\_particles,swarm\_epochs) \;    i $\leftarrow$ 0 \;    \ForEach{$ h_1 \in  hyperparameters$}{        \While{$PS.swarm\_epochs \ne 0$}{      $h_1 \leftarrow PS.maximize(nn.AUC,h_1 ) using algorithm 1$\;     }     $ nn.save\_opt\_hyperparam((h_1 )) $ \;    }    $ nn.train\_NN(training\_set()) $\;
	\end{algorithm}	
	
	In Algorithm \ref{alg:PDM_Opt} we depict a Particle Deep
	Framework (PDF) iteration. First the neural network is loaded with its
	pre-selected layers and number of neurons. Initially, the three hyperparameters
	batch size, number of epochs and learning rate {[}b,e,lr{]} are randomly
	initialized. Next a particle swarm comprised of a pre-selected number
	of particles ($n\_particles$) and number of iterations ($swarm\_epochs$)
	is generated. Then algorithm \ref{alg:PSO_Max} is utilized to identify
	the value of the hyperparameters that are being optimized and maximize
	the AUC value of the neural network ($h_1 \leftarrow PS.maximize(nn.AUC,h_1 )$).
	The process is repeated for every hyperparameter that is being optimized,
	the identified values of which are utilized to train the final neural
	network. 
	
	To validate the optimised deep MLP model, we selected the Bot-IoT 
	\cite{koroniotis2019towards}, as it is a contemporary dataset that 
	combines IoT and non-IoT traces and attacks. The dataset was 
	partitioned 80\% and 20\% sets for training and respectively. 
	The features of both training and testing sets were normalized with 
	min-max in a range between [0,1]. As there exists no standard procedure 
	for selecting optimal hyperparameters, at first, the values were manually 
	selected and the deep MLP model trained, after which we employed the 
	PDF as given by Algorithm \ref{alg:PDM_Opt}. 
	
	The performance of a neural network can be greatly affected by 
	the hyperparameters. In this study, we focus on optimising three, 
	learning rate, batch size and epochs. The learning rate is a decimal 
	between '0' and '1', that regulates the rate of change of the weights 
	during training. The batch size determines how many records are processed before the neural network's weights are updated, and 
	the number of epochs is the times that the entire dataset is 
	processed by the network during training. In the PDF, as is evident 
	in Algorithm \ref{alg:PDM_Opt}, each hyperparameter is optimised 
	separately, in order to reduce the search space that each PSO would 
	need to traverse. Otherwise, the search space would be equal to 
	\textit{Batch\_size\_Size * Epochs\_Size * Learning\_rate\_Size}.
	
	Because the deep model performed binary classification between 
	normal and attack traffic, the logistic cost function was selected 
	\cite{zhang2018generalized}, with its equation given below. In 
	addition, due to class imbalances in the Bot-IoT dataset, weights 
	were introduced in order to counter any detrimental effects they 
	might have had to the classifier's performances. 
	
	\begin{equation}   C=  -\frac{1}{m}\sum_{i=1}^{m}(y_i\log(\hat{y_i})+(1-y_i)\log(1-\hat{y_i}))  \end{equation}
	
	With \textit{m} being the number of records in each batch, 
	$y_i$ being the real value and $\hat{y_i}$ the estimated 
	value of the class feature of the $i^{th}$ record. 
	
	Introducing the new weights to counter the imbalanced data, 
	alter the logistic cost equation like so: $w_1 y_i log(\hat{y_i})+w_0 (1-y_i ) log(1-\hat{y_i})$, 
	where $w_0$ is the weight for normal and $w_1$ for attack 
	records. During a PSO iteration, a particle reaches a new 
	location and trains a version of the MLP, using its new location 
	as a hyperparameter, preserving values that improve the model's 
	performance. The process of training and testing a deep MLP model 
	using the PDF is given in Algorithm Algorithm \ref{alg:Train_Test_DPM}.
	\begin{algorithm}
		\caption{\label{alg:Train_Test_DPM} Steps for training and testing the proposed
			particle deep model}
		\KwData{}   $S ={0:'batch', 1:'epochs', 2:'learning_rate'}$\;   $Results ={'batch': -1, 'epochs': -1, 'learning_rate': -1}$\;   $NN =loadNN_structure()\#Hyperparameters\ that\ aren't\ trained$\;   $n\_p =6 \# number\ of\ particles$\;   $n\_e =4 \# number\ of\ epochs$\;   $Results =randomInitialState()$\;   $task = \textquoteleft maximize\_AUC \textquoteright$\;       \For{$k=0; k<=2; k++$}{       $\# Runs\ once\ for\ each\ hyperparameter\ to\ optimize$\;    $particles=generateParticles(n\_p, n\_e)$\;    $bestHyper=runPSO(particles, task, S[k], NN)$\;    $Results[S[k]]=bestHyper$\;   }   $\#trains\ a\ model\ with\ the\ identified\ hyperparameters$\;   $ trainedNN=trainNN(NN, Results)$\;    $testNN(trainedNN)$\;  
	\end{algorithm}
	As the AUC value is used to determine the wellness of a 
	hyperparameter, requiring the MLP to be trained beforehand, 
	the PDF execution time may be excessive, with recorded times 
	for our experiments and for each hyperparameter being: 4 hours 
	for batch, 3 hours for epochs and 4 for learning rate optimisation. 
	In addition, 7 minutes were required to train the final MLP model, 
	while its throughput was 14,762 records/second. The complexity 
	of the proposed PDF used to adjust one hyperparameter and for 
	each iteration is equal to $O(n_p*(mlp+1)$, with $n_p$ denoting 
	the number of particles spawned by the PSO and $mlp$ the time 
	complexity for training and testing an MLP model.
	
	\section{Results and Discussion}
	\subsection{Dataset and evaluation metrics}
		The contemporary Bot-IoT dataset \cite{koroniotis2019towards} was selected for training and testing the proposed PDF. The Bot-IoT combines IoT and non-IoT traffic, representing  a smart home deployment, with the former generated by using Node-Red \cite{2018i} and the entire dataset reaching 72.000.000 records and a total of 16.7 GB for its csv format. Specifically, we employed the "10-best feature" version of the Bot-IoT dataset, from which we utilised 2.934.817 records for training (80\%) and 733.705 records for testing (20\%). For evaluation purposes, we selected six metrics, accuracy, precision, recall, FPR, FNR, F-measure. Accuracy represents the fraction of correctly classified records $(TP+TN)/(TP+TN+FP+FN) $, precision is the fraction of predicted as "positive" records which were correctly identified $TP/(TP+FP)$. Recall is the fraction of records correctly identified as "positive" from all positive records $TP/(TP+FN)$, the FPR and FNR are the fractions fo records incorrectly classified as "positive" ($FP/(FP+TN)$) or "negative" ($FN/(FN+TP)$) respectively. Finally, the F-measure is a measure of a model's accuracy, produced by calculating the harmonic mean of the precision and recall values  $2TP/(2TP+FP+FN)$. The three main attack categories represented in the dataset are Denial of Service, information theft and information gathering attacks, with each category further specialized into subtypes as explained here \cite{koroniotis2019towards}. 
				
		Experiments were performed on a laptop equipped with 16 GB RAM, Intel Core i7-6700HQ CPU @2.6GHz, and the programming language used for designing and training the deep MLP model and utilising PSO for hyperparameter optimisation was python. The packages that were used in this python environment were as follows: for data pre-processing Numpy and Pandas, for implementing the deep MLP TensorFlow accessed through Keras and for defining and running the PSO hyperparameter optimisation, Optunity\cite{claesen2014hyperparameter}.

\subsection{Experiments}
This subsection describes the results obtained through experimentation and testing the Particle Deep Framework (PDF) by employing the evaluation metrics described in the previous subsection. The deep neural network architecture that was chosen is that of the deep MLP, as MLP networks have been shown to be simple, powerful and flexible models, having the ability to model non-linear relations in data \cite{abiodun2018state,mahdavinejad2018machine}. The hyperparameters and architectures of the neural networks that were evaluated are given in Table \ref{table:NeuralNetworksWereTrained}, with the activation function for the hidden layers and the output layer being "relu" and "sigmoid" respectively. For the MLP's weight initialization we used the \textit{glorot uniform}.

In addition to an optimised MLP, a feature-compressing method was tested, in order to investigate its effects on the deep model's performance. At first, the features are converted by using the Normal distribution's probability density function, then for each record, weights are applied, which are obtained by calculating the average of the correlation coefficient matrix of the features, finally combining the resulting values by adding and normalising them (min-max). As was already mentioned, we applied weights to compensate for the class imbalances of the Bot-IoT dataset. Specifically, we used "1" for attack records and "4500" for normal records. Furthermore, we specified fixed values for the random seeds used by the python implementation of MLP and PSO (Keras, TensorFLow and Optunity) to make the results reproducible. The results of our experiments, including the application of the PDF, are displayed in Table \ref{table:NeuralNetworksWereTrained}.

The three neural networks depicted in Table \ref{table:NeuralNetworksWereTrained} are further discussed here. First, (i) is an unoptimized neural network, trained on randomly chosen hyperparameters. Its high accuracy value is misleading, as its false-positive rate is just under 89\% which is achieved due to class imbalances. Second, (ii) is an MLP, that was trained by using the PDF and data that has been compressed by combining the features into one, as was previously described in this subsection. The justification for investigating a compression method for the features was to reduce the training time. Although the model's accuracy was slightly reduced to 95\%, compared to the unoptimised version of this MLP, its false-positive rate was considerably reduced to 8\% with its false-negative rate increased to 5\% respectively. Finally, (iii) is an MLP with optimised hyperparameters, that was trained and tested by using the PDF, on the original 13-feature Bot-IoT dataset. This MLP outperformed the other two versions, achieving false positive and negative rates close to '0', while maintaining an accuracy of 99.9\% and a precision of '1'.

A number of reasons can clarify why the 13-feature MLP (iii) outperformed the compressed, single-feature MLP (ii). The single-feature MLP was trained on data that compressed the input features, which may have resulted in considerable information loss. Furthermore, the 13-feature MLP (iii) incorporates 240 more trainable weights than the single-feature MLP (ii) between the input and first hidden layers, allowing (iii) to detect even more complex patterns in the data than (ii).

\begin{table}
	
	\caption{\label{table:NeuralNetworksWereTrained} Neural Networks that were
		trained.}
	
	\centering{}\begin{tabular}{|p{0.15\textwidth}|p{0.2\textwidth}|p{0.2\textwidth}|p{0.2\textwidth}|} 				\hline 			
		&(i) Unoptimized NN&(ii) Optimized NN with compressed input&(iii) Optimized NN with 13-features input\\\hline 				Neurons per layer&13, 20, 40, 60, 80, 40, 10, 1&1, 20, 40, 60, 80, 40, 10, 1&13, 20, 40, 60, 80, 40, 10, 1\\\hline 				Epochs&2&12&12\\\hline 				Batch size&350&3064&732\\\hline 				Learning rate&0.2&0.0015&0.0015\\\hline 				Accuracy&0.999&0.947&0.999\\\hline 				Precision&0.999&0.999&1\\\hline 				Recall&0.999&0.947&0.999\\\hline 				FPR&0.884&0.081&0\\\hline 				FNR&9.269*$10^{-5}$&0.052&9.541*$10^{-5}$\\\hline 				F-measure&0.999&0.973&0.999\\\hline 			\end{tabular}
\end{table}

\begin{figure}[H] 	\minipage{0.45\textwidth} 			\includegraphics[width=\textwidth]{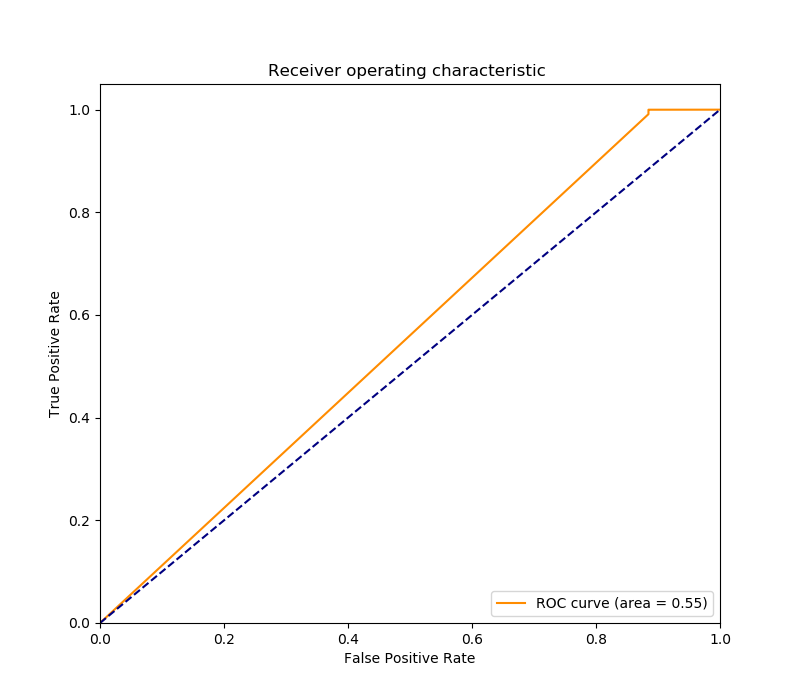} 			\caption{(i) Unoptimized NN} 		\endminipage \hfill 		\minipage{0.45\textwidth} 		\includegraphics[width=\textwidth]{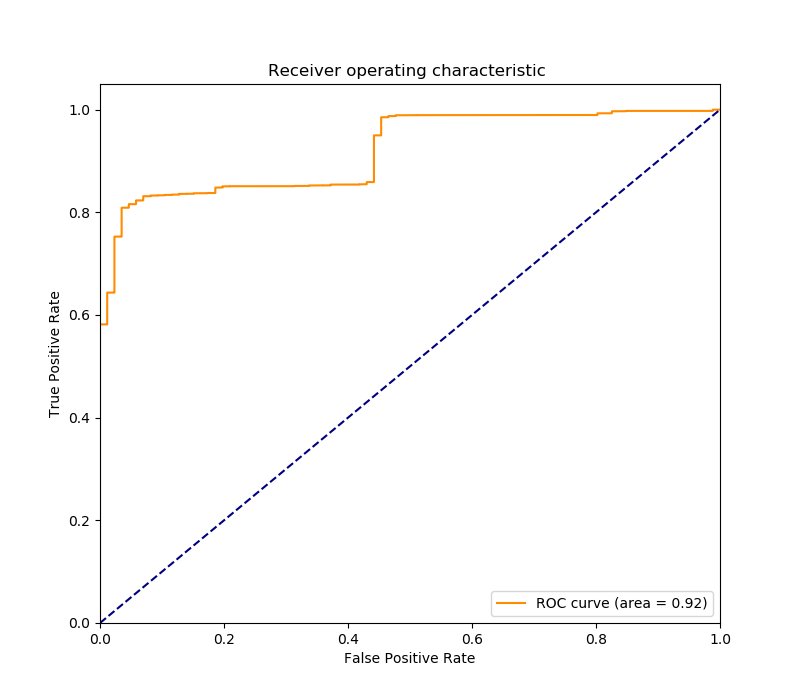} 		\caption{(ii) Optimized NN with compressed input} 		\endminipage \hfill  		\begin{center}\minipage{0.45\textwidth} 		\includegraphics[width=\textwidth]{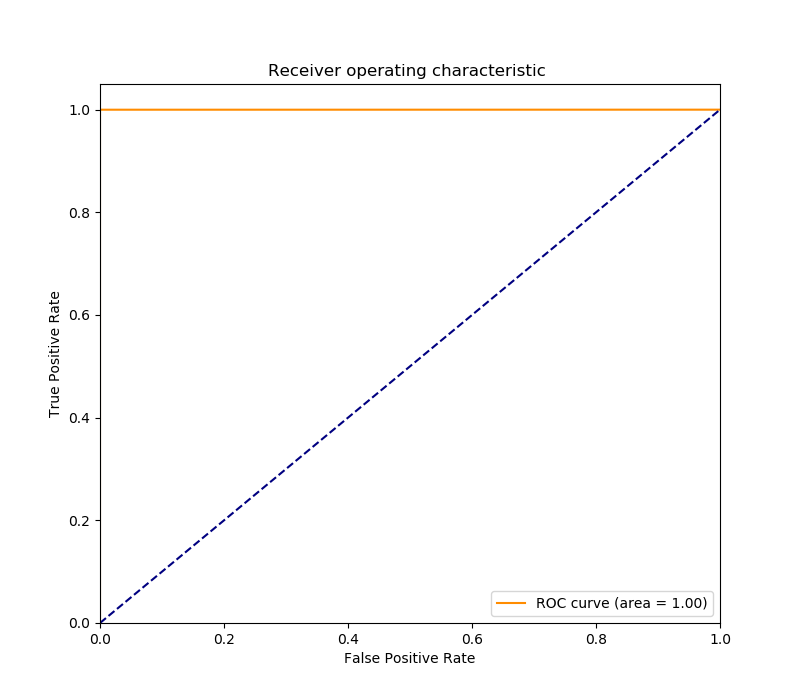} 		 		\endminipage \end{center}\caption{(iii) Optimized NN with 13-features input} \hfill 	
\end{figure}

\section{Advantages and limitations}

One advantage of the PDF is that it automates the hyperparameter tuning process, which originally was manual. Furthermore, the collection stage is carried out by reliable software that has seen wide use in industry. Additionally, the preservation stage is addressed, by using a cryptographic hashing algorithm, thus providing a mechanism to validate the integrity of data, the lack of which can cause a forensic investigation to be dismissed.

A limitation of the PDF, is that it requires considerable time to train a deep MLP model. This occurs because each particle trains an MLP on different hyperparameters, thus the size of the dataset, the number of layers and neurons of the MLP and the number of particles used in the swarm will increase time requirements. Furthermore, the PDF processes network flow data, thus any information in the body of collected packets is overlooked. In addition, countering spoofing attacks, that alter the IP address of the attacker, is challenging and may hinder an investigation.

\section{Conclusion}
	Due to the swift adoption of IoT systems by industry and the general public, attacks 
	targeting IoT networks have been increasing. This paper proposes a novel network 
	forensics framework Called Particle Deep Framework (PDF) for the detection and 
	analysis of cyber-attacks in IoT networks. First, the components of the PDF 
	and its correspondence to the forensic stages were explained. In its core, the 
	PDF combined Deep Learning as the base model and Particle Swarm Optimisation 
	for tuning its hyperparameters, with the contemporary Bot-IoT dataset 
	used to validate its performance. The PDF achieved very high attack detection 
	accuracy, at 99.9\%, with false positive and negative rates approaching zero, while 
	its classification speed was measured at 14,762 flows per second. As future work, 
	we intend to expand the PSO's functionality, by adjusting it to process multiple 
	hyperparameters, as-well-as test its effectiveness against other IoT deployments, like smart health networks.

	\section*{Acknowledgments}
	Nickolaos Koroniotis would like to thank the Commonwealth's support, which is provided to the aforementioned researcher in the form of an Australian Government Research Training Program Scholarship.
	\bibliographystyle{IEEEtran}
	\bibliography{Network_Forensics_as_an_applicaiton_of_Particle_Swarm_and_Deep_Learni}

\begin{thebibliography}{10}
\providecommand{\url}[1]{#1}
\csname url@samestyle\endcsname
\providecommand{\newblock}{\relax}
\providecommand{\bibinfo}[2]{#2}
\providecommand{\BIBentrySTDinterwordspacing}{\spaceskip=0pt\relax}
\providecommand{\BIBentryALTinterwordstretchfactor}{4}
\providecommand{\BIBentryALTinterwordspacing}{\spaceskip=\fontdimen2\font plus
\BIBentryALTinterwordstretchfactor\fontdimen3\font minus
  \fontdimen4\font\relax}
\providecommand{\BIBforeignlanguage}[2]{{%
\expandafter\ifx\csname l@#1\endcsname\relax
\typeout{** WARNING: IEEEtran.bst: No hyphenation pattern has been}%
\typeout{** loaded for the language `#1'. Using the pattern for}%
\typeout{** the default language instead.}%
\else
\language=\csname l@#1\endcsname
\fi
#2}}
\providecommand{\BIBdecl}{\relax}
\BIBdecl

\bibitem{Cook2019}
\BIBentryALTinterwordspacing
S.~Cook, ``60+ iot statistics and facts,'' \emph{Comparitech}, 2019. [Online].
  Available:
  \url{https://www.comparitech.com/internet-providers/iot-statistics/}
\BIBentrySTDinterwordspacing

\bibitem{Corporation2019}
\BIBentryALTinterwordspacing
S.~Corporation, ``Internet security threat report, volume 24,'' Symantec, Tech.
  Rep.~24, 2019. [Online]. Available:
  \url{https://www.symantec.com/content/dam/symantec/docs/reports/istr-24-2019-en.pdf}
\BIBentrySTDinterwordspacing

\bibitem{koroniotis2019forensics}
N.~Koroniotis, N.~Moustafa, and E.~Sitnikova, ``Forensics and deep learning
  mechanisms for botnets in internet of things: A survey of challenges and
  solutions,'' \emph{IEEE Access}, vol.~7, pp. 61\,764--61\,785, 2019.

\bibitem{ali2018cyber}
B.~Ali and A.~Awad, ``Cyber and physical security vulnerability assessment for
  iot-based smart homes,'' \emph{Sensors}, vol.~18, no.~3, p. 817, 2018.

\bibitem{Robberts2019}
C.~Robberts and J.~Toft, ``Finding vulnerabilities in iot devices: Ethical
  hacking of electronic locks,'' 2019.

\bibitem{conti2018internet}
M.~Conti, A.~Dehghantanha, K.~Franke, and S.~Watson, ``Internet of things
  security and forensics: Challenges and opportunities,'' \emph{Future
  Generation Computer Systems}, vol.~78, pp. 544--546, 2018.

\bibitem{ronen2017iot}
E.~Ronen, A.~Shamir, A.-O. Weingarten, and C.~O'Flynn, ``Iot goes nuclear:
  Creating a zigbee chain reaction,'' in \emph{2017 IEEE Symposium on Security
  and Privacy (SP)}.\hskip 1em plus 0.5em minus 0.4em\relax IEEE, 2017, pp.
  195--212.

\bibitem{meffert2017forensic}
C.~Meffert, D.~Clark, I.~Baggili, and F.~Breitinger, ``Forensic state
  acquisition from internet of things (fsaiot): A general framework and
  practical approach for iot forensics through iot device state acquisition,''
  in \emph{Proceedings of the 12th International Conference on Availability,
  Reliability and Security}.\hskip 1em plus 0.5em minus 0.4em\relax ACM, 2017,
  p.~56.

\bibitem{al2017internet}
S.~Al-Sarawi, M.~Anbar, K.~Alieyan, and M.~Alzubaidi, ``Internet of things
  (iot) communication protocols,'' in \emph{2017 8th International conference
  on information technology (ICIT)}.\hskip 1em plus 0.5em minus 0.4em\relax
  IEEE, 2017, pp. 685--690.

\bibitem{kaur2012digital}
R.~Kaur and A.~Kaur, ``Digital forensics,'' \emph{International Journal of
  Computer Applications}, vol.~50, no.~5, 2012.

\bibitem{hossain2018fif}
M.~Hossain, Y.~Karim, and R.~Hasan, ``Fif-iot: A forensic investigation
  framework for iot using a public digital ledger,'' in \emph{2018 IEEE
  International Congress on Internet of Things (ICIOT)}.\hskip 1em plus 0.5em
  minus 0.4em\relax IEEE, 2018, pp. 33--40.

\bibitem{hossain2018probe}
M.~M. Hossain, R.~Hasan, and S.~Zawoad, ``Probe-iot: A public digital ledger
  based forensic investigation framework for iot.'' in \emph{INFOCOM
  Workshops}, 2018, pp. 1--2.

\bibitem{Cebe2018}
M.~{Cebe}, E.~{Erdin}, K.~{Akkaya}, H.~{Aksu}, and S.~{Uluagac},
  ``Block4forensic: An integrated lightweight blockchain framework for
  forensics applications of connected vehicles,'' \emph{IEEE Communications
  Magazine}, vol.~56, no.~10, pp. 50--57, OCTOBER 2018.

\bibitem{shone2018deep}
N.~Shone, T.~N. Ngoc, V.~D. Phai, and Q.~Shi, ``A deep learning approach to
  network intrusion detection,'' \emph{IEEE Transactions on Emerging Topics in
  Computational Intelligence}, vol.~2, no.~1, pp. 41--50, 2018.

\bibitem{prabakaran2018survey}
S.~Prabakaran and S.~Mitra, ``Survey of analysis of crime detection techniques
  using data mining and machine learning,'' in \emph{Journal of Physics:
  Conference Series}, vol. 1000, no.~1.\hskip 1em plus 0.5em minus 0.4em\relax
  IOP Publishing, 2018, p. 012046.

\bibitem{wang2015applications}
Z.~Wang, ``The applications of deep learning on traffic identification,''
  \emph{BlackHat USA}, vol.~24, 2015.

\bibitem{zhao2017intrusion}
G.~Zhao, C.~Zhang, and L.~Zheng, ``Intrusion detection using deep belief
  network and probabilistic neural network,'' in \emph{2017 IEEE International
  Conference on Computational Science and Engineering (CSE) and IEEE
  International Conference on Embedded and Ubiquitous Computing (EUC)},
  vol.~1.\hskip 1em plus 0.5em minus 0.4em\relax IEEE, 2017, pp. 639--642.

\bibitem{Zela2018}
A.~Zela, A.~Klein, S.~Falkner, and F.~Hutter, ``Towards automated deep
  learning: Efficient joint neural architecture and hyperparameter search,'' in
  \emph{ICML 2018 AutoML Workshop}, Jul. 2018.

\bibitem{wang2018stealing}
B.~Wang and N.~Z. Gong, ``Stealing hyperparameters in machine learning,'' in
  \emph{2018 IEEE Symposium on Security and Privacy (SP)}.\hskip 1em plus 0.5em
  minus 0.4em\relax IEEE, 2018, pp. 36--52.

\bibitem{chen2018learning}
T.~Chen, L.~Zheng, E.~Yan, Z.~Jiang, T.~Moreau, L.~Ceze, C.~Guestrin, and
  A.~Krishnamurthy, ``Learning to optimize tensor programs,'' in \emph{Advances
  in Neural Information Processing Systems}, 2018, pp. 3389--3400.

\bibitem{wang2018combination}
J.~Wang, J.~Xu, and X.~Wang, ``Combination of hyperband and bayesian
  optimization for hyperparameter optimization in deep learning,'' \emph{arXiv
  preprint arXiv:1801.01596}, 2018.

\bibitem{stamoulis2018hyperpower}
D.~Stamoulis, E.~Cai, D.-C. Juan, and D.~Marculescu, ``Hyperpower: Power-and
  memory-constrained hyper-parameter optimization for neural networks,'' in
  \emph{2018 Design, Automation \& Test in Europe Conference \& Exhibition
  (DATE)}.\hskip 1em plus 0.5em minus 0.4em\relax IEEE, 2018, pp. 19--24.

\bibitem{watson2016digital}
S.~Watson and A.~Dehghantanha, ``Digital forensics: the missing piece of the
  internet of things promise,'' \emph{Computer Fraud \& Security}, vol. 2016,
  no.~6, pp. 5--8, 2016.

\bibitem{chernyshev2018internet}
M.~Chernyshev, S.~Zeadally, Z.~Baig, and A.~Woodward, ``Internet of things
  forensics: The need, process models, and open issues,'' \emph{IT
  Professional}, vol.~20, no.~3, pp. 40--49, 2018.

\bibitem{palmer2001road}
G.~L. Palmer, ``A road map for digital forensics research-report from the first
  digital forensics research workshop (dfrws)(technical report dtr-t001-01
  final),'' \emph{Air Force Research Laboratory, Rome Research Site, Utica},
  pp. 1--48, 2001.

\bibitem{joseph2019analysis}
D.~P. Joseph and J.~Norman, ``An analysis of digital forensics in cyber
  security,'' in \emph{First International Conference on Artificial
  Intelligence and Cognitive Computing}.\hskip 1em plus 0.5em minus 0.4em\relax
  Springer, 2019, pp. 701--708.

\bibitem{le2018biff}
D.-P. Le, H.~Meng, L.~Su, S.~L. Yeo, and V.~Thing, ``Biff: A blockchain-based
  iot forensics framework with identity privacy,'' in \emph{TENCON 2018-2018
  IEEE Region 10 Conference}.\hskip 1em plus 0.5em minus 0.4em\relax IEEE,
  2018, pp. 2372--2377.

\bibitem{valjarevic2015comprehensive}
A.~Valjarevic and H.~S. Venter, ``A comprehensive and harmonized digital
  forensic investigation process model,'' \emph{Journal of forensic sciences},
  vol.~60, no.~6, pp. 1467--1483, 2015.

\bibitem{caviglione2017future}
L.~Caviglione, S.~Wendzel, and W.~Mazurczyk, ``The future of digital forensics:
  Challenges and the road ahead,'' \emph{IEEE Security \& Privacy}, vol.~15,
  no.~6, pp. 12--17, 2017.

\bibitem{han2016survey}
S.-W. Han, H.~Kwon, C.~Hahn, D.~Koo, and J.~Hur, ``A survey on mitm and its
  countermeasures in the tls handshake protocol,'' in \emph{2016 Eighth
  International Conference on Ubiquitous and Future Networks (ICUFN)}.\hskip
  1em plus 0.5em minus 0.4em\relax IEEE, 2016, pp. 724--729.

\bibitem{liao2009network}
N.~Liao, S.~Tian, and T.~Wang, ``Network forensics based on fuzzy logic and
  expert system,'' \emph{Computer Communications}, vol.~32, no.~17, pp.
  1881--1892, 2009.

\bibitem{ahmed2018sairf}
A.~A. Ahmed and M.~F. Mohammed, ``Sairf: A similarity approach for attack
  intention recognition using fuzzy min-max neural network,'' \emph{Journal of
  Computational Science}, vol.~25, pp. 467--473, 2018.

\bibitem{yudhana2018ddos}
A.~Yudhana, I.~Riadi, and F.~Ridho, ``Ddos classification using neural network
  and na{\"\i}ve bayes methods for network forensics,'' \emph{INTERNATIONAL
  JOURNAL OF ADVANCED COMPUTER SCIENCE AND APPLICATIONS}, vol.~9, no.~11, pp.
  177--183, 2018.

\bibitem{nguyen2014approach}
K.~Nguyen, D.~Tran, W.~Ma, and D.~Sharma, ``An approach to detect network
  attacks applied for network forensics,'' in \emph{2014 11th International
  Conference on Fuzzy Systems and Knowledge Discovery (FSKD)}.\hskip 1em plus
  0.5em minus 0.4em\relax IEEE, 2014, pp. 655--660.

\bibitem{alrawashdeh2016toward}
K.~Alrawashdeh and C.~Purdy, ``Toward an online anomaly intrusion detection
  system based on deep learning,'' in \emph{2016 15th IEEE International
  Conference on Machine Learning and Applications (ICMLA)}.\hskip 1em plus
  0.5em minus 0.4em\relax IEEE, 2016, pp. 195--200.

\bibitem{azmoodeh2018robust}
A.~Azmoodeh, A.~Dehghantanha, and K.-K.~R. Choo, ``Robust malware detection for
  internet of (battlefield) things devices using deep eigenspace learning,''
  \emph{IEEE Transactions on Sustainable Computing}, vol.~4, no.~1, pp. 88--95,
  2018.

\bibitem{lecun2015deep}
Y.~LeCun, Y.~Bengio, and G.~Hinton, ``Deep learning,'' \emph{nature}, vol. 521,
  no. 7553, p. 436, 2015.

\bibitem{brun2018deep}
O.~Brun, Y.~Yin, and E.~Gelenbe, ``Deep learning with dense random neural
  network for detecting attacks against iot-connected home environments,''
  \emph{Procedia computer science}, vol. 134, pp. 458--463, 2018.

\bibitem{pektacs2018botnet}
A.~Pekta{\c{s}} and T.~Acarman, ``Botnet detection based on network flow
  summary and deep learning,'' \emph{International Journal of Network
  Management}, p. e2039, 2018.

\bibitem{cheng2018d2pi}
R.~Cheng and G.~Watson, ``D2pi: Identifying malware through deep packet
  inspection with deep learning,'' 2018.

\bibitem{alrashdi2019ad}
I.~Alrashdi, A.~Alqazzaz, E.~Aloufi, R.~Alharthi, M.~Zohdy, and H.~Ming,
  ``Ad-iot: anomaly detection of iot cyberattacks in smart city using machine
  learning,'' in \emph{2019 IEEE 9th Annual Computing and Communication
  Workshop and Conference (CCWC)}.\hskip 1em plus 0.5em minus 0.4em\relax IEEE,
  2019, pp. 0305--0310.

\bibitem{babun2018iotdots}
L.~Babun, A.~K. Sikder, A.~Acar, and A.~S. Uluagac, ``Iotdots: A digital
  forensics framework for smart environments,'' \emph{arXiv preprint
  arXiv:1809.00745}, 2018.

\bibitem{kennedy1995particle}
J.~Kennedy and R.~Eberhart, ``Particle swarm optimization (pso),'' in
  \emph{Proc. IEEE International Conference on Neural Networks, Perth,
  Australia}, 1995, pp. 1942--1948.

\bibitem{marini2015particle}
F.~Marini and B.~Walczak, ``Particle swarm optimization (pso). a tutorial,''
  \emph{Chemometrics and Intelligent Laboratory Systems}, vol. 149, pp.
  153--165, 2015.

\bibitem{wang2018particle}
D.~Wang, D.~Tan, and L.~Liu, ``Particle swarm optimization algorithm: an
  overview,'' \emph{Soft Computing}, vol.~22, no.~2, pp. 387--408, 2018.

\bibitem{shi1998modified}
Y.~Shi and R.~Eberhart, ``A modified particle swarm optimizer,'' in \emph{1998
  IEEE international conference on evolutionary computation proceedings. IEEE
  world congress on computational intelligence (Cat. No. 98TH8360)}.\hskip 1em
  plus 0.5em minus 0.4em\relax IEEE, 1998, pp. 69--73.

\bibitem{eberhart2001tracking}
R.~C. Eberhart and Y.~Shi, ``Tracking and optimizing dynamic systems with
  particle swarms,'' in \emph{Proceedings of the 2001 Congress on Evolutionary
  Computation (IEEE Cat. No. 01TH8546)}, vol.~1.\hskip 1em plus 0.5em minus
  0.4em\relax IEEE, 2001, pp. 94--100.

\bibitem{eberhart1996computational}
R.~Eberhart, P.~Simpson, and R.~Dobbins, \emph{Computational intelligence PC
  tools}.\hskip 1em plus 0.5em minus 0.4em\relax Academic Press Professional,
  Inc., 1996.

\bibitem{clerc1999swarm}
M.~Clerc, ``The swarm and the queen: towards a deterministic and adaptive
  particle swarm optimization,'' in \emph{Proceedings of the 1999 congress on
  evolutionary computation-CEC99 (Cat. No. 99TH8406)}, vol.~3.\hskip 1em plus
  0.5em minus 0.4em\relax IEEE, 1999, pp. 1951--1957.

\bibitem{eberhart2000comparing}
R.~C. Eberhart and Y.~Shi, ``Comparing inertia weights and constriction factors
  in particle swarm optimization,'' in \emph{Proceedings of the 2000 congress
  on evolutionary computation. CEC00 (Cat. No. 00TH8512)}, vol.~1.\hskip 1em
  plus 0.5em minus 0.4em\relax IEEE, 2000, pp. 84--88.

\bibitem{kennedy1997discrete}
J.~Kennedy and R.~C. Eberhart, ``A discrete binary version of the particle
  swarm algorithm,'' in \emph{1997 IEEE International conference on systems,
  man, and cybernetics. Computational cybernetics and simulation},
  vol.~5.\hskip 1em plus 0.5em minus 0.4em\relax IEEE, 1997, pp. 4104--4108.

\bibitem{van2004cooperative}
F.~Van~den Bergh and A.~P. Engelbrecht, ``A cooperative approach to particle
  swarm optimization,'' \emph{IEEE transactions on evolutionary computation},
  vol.~8, no.~3, pp. 225--239, 2004.

\bibitem{mendes2004fully}
R.~Mendes, J.~Kennedy, and J.~Neves, ``The fully informed particle swarm:
  simpler, maybe better,'' \emph{IEEE transactions on evolutionary
  computation}, vol.~8, no.~3, pp. 204--210, 2004.

\bibitem{2018Wireshark}
\BIBentryALTinterwordspacing
Wireshark tool. [Online]. Available: \url{https://www.wireshark.org/}
\BIBentrySTDinterwordspacing

\bibitem{2018Tcpdump}
\BIBentryALTinterwordspacing
Tcpdump tool. [Online]. Available: \url{https://www.tcpdump.org}
\BIBentrySTDinterwordspacing

\bibitem{2018Ettercap}
\BIBentryALTinterwordspacing
Ettercap tool. [Online]. Available: \url{https://www.ettercap-project.org/}
\BIBentrySTDinterwordspacing

\bibitem{koroniotis2019towards}
N.~Koroniotis, N.~Moustafa, E.~Sitnikova, and B.~Turnbull, ``Towards the
  development of realistic botnet dataset in the internet of things for network
  forensic analytics: Bot-iot dataset,'' \emph{Future Generation Computer
  Systems}, 2019.

\bibitem{2018p}
\BIBentryALTinterwordspacing
Tshark network analysis tool. [Online]. Available:
  \url{https://www.wireshark.org/}
\BIBentrySTDinterwordspacing

\bibitem{armknecht2015privacy}
F.~Armknecht and A.~Dewald, ``Privacy-preserving email forensics,''
  \emph{Digital Investigation}, vol.~14, pp. S127--S136, 2015.

\bibitem{2018j}
\BIBentryALTinterwordspacing
Argus tool. [Online]. Available: \url{https://qosient.com/argus/index.shtml}
\BIBentrySTDinterwordspacing

\bibitem{2018a}
\BIBentryALTinterwordspacing
Bro tool. [Online]. Available: \url{https://www.zeek.org/}
\BIBentrySTDinterwordspacing

\bibitem{lu2015fuzzy}
X.~Lu and M.~Liu, ``A fuzzy logic controller tuned with pso for delta robot
  trajectory control,'' in \emph{IECON 2015-41st Annual Conference of the IEEE
  Industrial Electronics Society}.\hskip 1em plus 0.5em minus 0.4em\relax IEEE,
  2015, pp. 004\,345--004\,351.

\bibitem{Couceiro2016}
M.~Couceiro and P.~Ghamisi, \emph{Particle Swarm Optimization}.\hskip 1em plus
  0.5em minus 0.4em\relax Cham: Springer International Publishing, 2016, pp.
  1--10.

\bibitem{bonyadi2017particle}
M.~R. Bonyadi and Z.~Michalewicz, ``Particle swarm optimization for single
  objective continuous space problems: a review,'' 2017.

\bibitem{ab2015comprehensive}
M.~N. Ab~Wahab, S.~Nefti-Meziani, and A.~Atyabi, ``A comprehensive review of
  swarm optimization algorithms,'' \emph{PloS one}, vol.~10, no.~5, p.
  e0122827, 2015.

\bibitem{zhang2018generalized}
Z.~Zhang and M.~Sabuncu, ``Generalized cross entropy loss for training deep
  neural networks with noisy labels,'' in \emph{Advances in neural information
  processing systems}, 2018, pp. 8778--8788.

\bibitem{2018i}
\BIBentryALTinterwordspacing
Node-red tool. [Online]. Available: \url{https://nodered.org/}
\BIBentrySTDinterwordspacing

\bibitem{claesen2014hyperparameter}
M.~Claesen, J.~Simm, D.~Popovic, and B.~Moor, ``Hyperparameter tuning in python
  using optunity,'' in \emph{Proceedings of the International Workshop on
  Technical Computing for Machine Learning and Mathematical Engineering},
  vol.~1, 2014, p.~3.

\bibitem{abiodun2018state}
O.~I. Abiodun, A.~Jantan, A.~E. Omolara, K.~V. Dada, N.~A. Mohamed, and
  H.~Arshad, ``State-of-the-art in artificial neural network applications: A
  survey,'' \emph{Heliyon}, vol.~4, no.~11, p. e00938, 2018.

\bibitem{mahdavinejad2018machine}
M.~S. Mahdavinejad, M.~Rezvan, M.~Barekatain, P.~Adibi, P.~Barnaghi, and A.~P.
  Sheth, ``Machine learning for internet of things data analysis: A survey,''
  \emph{Digital Communications and Networks}, vol.~4, no.~3, pp. 161--175,
  2018.

\end{thebibliography}
	
	\vspace{2cm}

	\section*{Authors}
	\noindent \textbf{Nickolaos Koroniotis}  is a PhD student at UNSW Canberra. He received his Bachelors in Informatics and Telematics in 2014 and his Masters in Web Engineering and Applications in 2016. He enrolled in UNSW Canberra to initiate his PhD studies in February 2017 in the field of Cyber security with a particular interest in Network Forensics and the IoT. \\
	
		\noindent \textbf{Nour Moustafa} $(SM19)$   is a Lecturer and Theme Lead of Intelligent Security at the School of Engineering and Information Technology, the University of New South Wales, Canberra, Australia.  He was a Postdoctoral Fellow at UNSW Canberra from June 2017 till December 2018. He received his PhD degree in the field of Cyber Security from UNSW Canberra in 2017. He obtained his Bachelor and master's degree in computer science in 2009 and 2014, respectively, from Helwan University, Egypt. His areas of interests include Cyber Security, in particular, network security, big data analytics, service orchestration, host- and network- intrusion detection systems, statistics, and machine/deep learning algorithms. He is interested in designing and developing threat detection and forensic mechanisms to the Industry 4.0 technology for identifying malicious activities from cloud computing, fog computing, IoT and industrial control systems. He services as a guest associate editor at IEEE Access and a reviewer of many high-tier journals and conferences in the domains of security and computing.

\end{document}